# Arsenic diffusion in MOVPE-Grown GaAs/Ge epitaxial structures

V. Orejuela*, I. Rey-Stolle, I. García*
Universidad Politécnica de Madrid
Instituto de Energía Solar
Semiconductores III-V
Avenida Complutense, 30
Madrid 28040, Spain
*Corresponding authors' emails: victor.orejuela@ies.upm.es , igarcia@ies.upm.es

E. García-Tabares
Universidad Carlos III de Madrid
Edificio Torres Quevedo
Departamento de Física
Leganés (Madrid) 28911, Spain

ORCID IDs:
V. Orejuela: https://orcid.org/0000-0003-3361-6509
I. Rey-Stolle: https://orcid.org/0000-0002-4919-5609
I. García: https://orcid.org/0000-0002-9895-2020
E. García-Tabares: https://orcid.org/0000-0002-4080-376X

**Germanium is reemerging as a prominent material in the semiconductor field, particularly for electronic applications, photonics, photovoltaics and thermophotovoltaics. Its combination with III-V compound semiconductors through epitaxial growth by metal organic vapor phase epitaxy (MOVPE) is instrumental and thus, the comprehension of the sequential stages in such epitaxial processes is of great importance. During the deposition of GaAs on p-type Ge, the formation of n/p junctions occurs when As diffuses into Ge. It is found that this formation begins in the so-called $AsH_3$ preexposure where Ge substrate is firstly exposed to $AsH_3$. Also important is the fact that both free carrier profiles and As profiles indicate that prolonged $AsH_3$ preexposure times lead to deeper diffusion depths for the same process time. This effect is concomitant with the degradation of the Ge surface morphology, characterized by higher roughness as the $AsH_3$ preexposure duration is extended. Contrary to ion-implanted As in germanium, which shows quadratic dependent diffusivity, our MOVPE investigation using $AsH_3$ indicates a linear relationship, consistent with Takenaka et al.'s MOVPE study using TBAs. Analyzing As profiles alongside simulations, with and without subsequent GaAs epitaxy, suggests the generation of Ge vacancies during the process, contributing to deeper As diffusion.**

## 1. Introduction

The low bandgap of germanium (Ge) makes it a suited material for devices such as near-infrared photodetectors or single-junction thermophotovoltaic (TPV) receivers.[1], [2], [3], [4], [5] Semiconductor devices combining Ge and III-V materials are typically implemented in the field of photovoltaics, where Ge is used as substrate and bottom cell in high efficiency multijunction solar cells.[6] Metal-Organic Vapor Phase Epitaxy (MOVPE) is the most widespread method for the growth of III-V semiconductor structures, at an industrial level. Ge-based photovoltaic device structures, such as the aforementioned multijunction solar cells, are obtained by the epitaxial growth of III-V layers on Ge substrates. The high temperatures involved in the process cause a variety of effects including segregation and cross-diffusion of III-V elements and Ge.[7] In fact, this effect allows the formation of pn junctions in Ge. For example, in photovoltaics a p-type Ge substrate is used and a lattice-matched GaAs or GaInP layer is deposited on top to passivate the Ge surface and minimize the surface minority carrier recombination. A n/p junction is created at the same time by diffusion of As or P, which are n-type dopants, into p-type Ge. GaInP is typically preferred because P diffusion is slower than As diffusion.[8], [9], [10], [11] However, GaAs offers a potentially lower deposition cost, since only two sources are needed, and no precise control over composition is required as a small lattice mismatch (0.07%) is produced on Ge. The faster diffusion of As might not be problematic for device structures not subjected to strong thermal loads after the deposition of the III-V layer, such as single-junction Ge photovoltaic cells. In fact, high performance TPV receivers using GaAs/Ge single junction cells (SJC) have already been demonstrated.[5], [12]

The diffusion of As into Ge is an essential step in the formation of the n/p junction in these photovoltaic devices. The final diffusion profile determines their performance. Therefore, a deep understanding of this diffusion process is required for the design of MOVPE deposition processes. However, many of experimental data reported in the literature about diffusion of As or P in Ge are obtained by applying a thermal load to ionically implanted Ge with these elements or to Ge-As alloys on pure Ge.[13], [14], [15], [16] Multiple studies of As diffusion during a MOVPE process have focused on the reduction of GaAs/Ge cross-diffusion by low temperature GaAs deposition or by growing a thin layer of AlAs on Ge.[7], [17] Nevertheless, studies of the evolution and impact of As diffusion during epitaxial processes have not thoroughly been conducted. The diffusion is affected by the MOVPE ambient and growth parameters, resulting in an intricate combination of cross-related effects. As an example, for a fixed thermal load, drastically different As diffusion profiles are observed depending on the pre-treatment of the surface or the deposition conditions of the GaAs layer.

In this work we contribute to elucidate how the main growth parameters involved in the deposition of GaAs/Ge structures affect the As diffusion profiles in Ge. The objective is to improve the control over the MOVPE process in achieving the desired dopant profiles.

## 2. Experimental

The experiments were conducted in a horizontal, research-scale MOVPE reactor (AIX200/4) featuring an infrared-lamp heater. The reactor pressure is kept at 100 mbar except during the GaAs deposition steps, when the pressure is increased to 250 mbar to raise the partial pressure of $AsH_3$, as it was found to be beneficial for the morphology of the nucleated GaAs layers. All samples were grown on (100) Ge wafers with a 6⁰ offcut towards the nearest (111) plane with low dopings of Ga (p-type with a resistivity of 1.5-4.5 Ohm·cm). The Ge wafer surfaces are covered by a thin oxide layer that is removed by a thermal treatment at the beginning of the deposition process. The surface condition is monitored by in-situ measurements of the Reflectance Anisotropy Spectroscopy (RAS) using an EpiRas 2000 tool. The precursors were high purity arsine ($AsH_3$) and trimethylgallium (TMGa). During the growth, molar flows of $4.4·10^{-5}$ mol·min$^{-1}$ and $2·10^{-2}$ mol·min$^{-1}$ were used for TMGa and $AsH_3$, respectively. For the $AsH_3$ pre-exposure stage (stage 3), the molar flow was $2·10^{-2}$ mol·min$^{-1}$. Total flow rates for these configurations remained constant at 5000 sccm during $AsH_3$ preexposure and 14000 sccm during GaAs growth. During the cool-down (see **Figure 1**), the molar flow of $AsH_3$ is kept at $2.2·10^{-3}$ mol·min$^{-1}$ to prevent As desorption until 300 ºC, when As desorption is negligible.[18] Di-tert-butylsilane (DTBSi) was employed to n-type dope the GaAs layer with Si. The carrier gas was high purity $H_2$. A growth temperature of 640 ºC and a growth rate of 0.75 µm·h$^{-1}$ were kept constant for every sample. To ensure experimental consistency, all epitaxy runs where performed with the same preconditioning of the reactor chamber, i.e., with a GaAs layer coating it.

Free charge carrier profiles were acquired by electrochemical capacitance-voltage (ECV) measurements utilizing a WEP CVP 21 tool and Tiron electrolyte. Secondary ion mass spectroscopy (SIMS) was performed on a SIMS Cameca IMS 4f-E6 analyzer (Cs$^+$, 14.5 keV), from which the chemical profiles of As concentrations were collected for key samples. The background of the SIMS profiles (roughly $2·10^{17}$ cm$^{-3}$) appears due to the difficulty of the mass spectrometer to separate the signals of $^{75}$As and $^{74}$Ge$^{1}$H. The surface of selected samples was analyzed using atomic force microscopy (AFM) by means of a Park XE-10 AFM Microscope, from which root mean square (RMS) roughness values were calculated.

Figure 1 shows a schematic profile of the temperature evolution in the MOVPE growths, which can be divided into five stages. Initially, at stage 1, the p-type Ge substrate has oxides on the surface, as mentioned before. As the temperature reaches values above 560 °C, the oxides start to desorb and As coming from the reactor chamber starts to dimerize on the Ge surface, which can be detected in the RAS signatures measured.[19] Then, the process continues with a gradual temperature ramp-up and stabilization at 640 °C. In the bake phase (stage 2), the Ge wafer is kept at this temperature for 5 min in order to fully remove the Ge oxides and to ensure that any possible foreign adsorbed elements are desorbed. Then AsH$_3$ is flown into the reactor and a monolayer of As covers the Ge surface (stage 3), stabilizing the surface.[20] However, the AsH$_3$ exposure initiates the diffusion of As into Ge and causes some surface roughening. In stage 4, GaAs deposition starts by flowing TMGa into the reactor chamber. During GaAs growth, cross diffusion of As and Ga into the Ge and vice versa take place across the GaAs/Ge interface. Meanwhile all elements continuously diffuse over time because of the thermal load. Although the cross-diffusion process involves multiple atoms (Ga, Ge, As and Si as n-type dopant in GaAs), this study primarily centers on the analysis of As diffusion. Regarding Ga, its influence was indirectly detected in the ECV profiles, as we will discuss later. In the final step, the GaAs layer deposition is complete (stage 5) and, the heater is turned off to start the cooldown stage. To prevent any surface damage, the As desorption is compensated by a small AsH$_3$ flow maintained down to 300 °C. The cooldown process is identical in all the samples studied in this work.

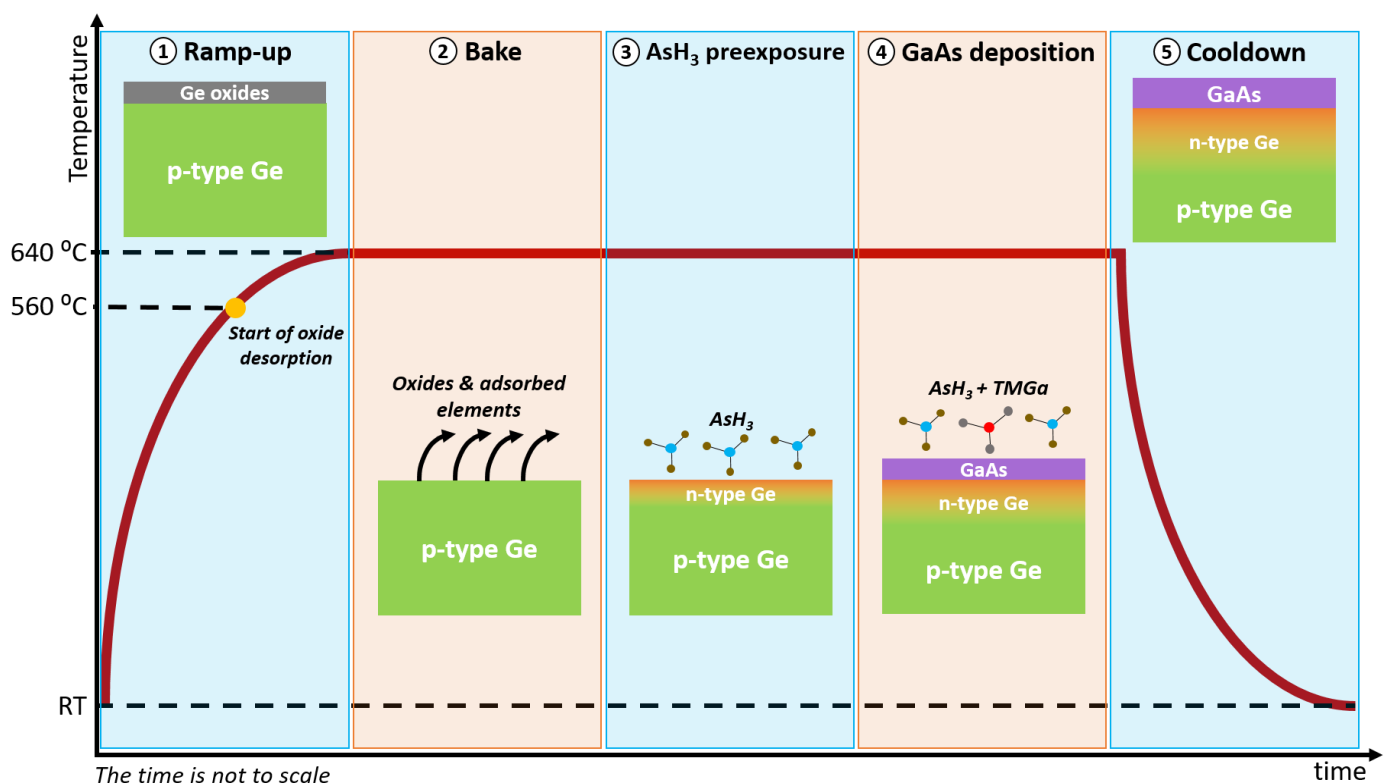

**Figure 1.** Schematic graph of the temperature profile in a GaAs deposition on Ge depicting the five stages of the process. Simplified illustrations of every stage are pictured. The time axis is not to scale.

## 3. Results

In this work, we aim to determine the influence on As diffusion of two key steps in MOVPE GaAs/Ge heteroepitaxy, namely, AsH$_3$ preexposure and GaAs deposition (stages 3 and 4 in Figure 1). These steps are the most relevant amid those listed in Figure 1 and correspond to the substrate preparation and the growth itself. A first issue to clarify is why preexposure is needed at all. The fact of the matter is that the growth of a GaAs layer on a Ge substrate is a heteroepitaxial process that involves the growth of a polar semiconductor (GaAs) on a non-polar substrate (Ge). Such growth entails a number of difficulties, being particularly important the avoidance of anti-phase domains (APD).[21] Anti-phase disorder can be minimized if GaAs growth proceeds on a Ge surface, covered in As dimers showing a single dimer orientation and double steps.[22] The preexposure step aims at achieving such surface by exposing the wafer to the group-V precursor for a short time before the growth starts.[23] The optimization of this step depends on reactor configuration, precursor partial pressure, and growth temperature but it is typically lower than a minute.

In an initial series of experiments, Ge substrates were exposed to varying preexposure times ($t_{Preexp}$) before the epitaxial deposition of 50-nm GaAs layers for 240 s on every sample. **Figure 2**(a) shows ECV profiles of this batch. The dips at the first nanometers (until about 25 nm) are due to the presence of Ga, a p-type dopant that compensates the n-type doping of the already-diffused As. Because of its lower diffusion coefficient relative to As, the Ga diffusion depth is much shallower than that of As.[24] This dip is followed by a plateau in the electron concentration profile, which finally drops at some depth into the Ge wafer. As it can be observed, the free carrier concentration corresponding to the plateaus decreases as $t_{Preexp}$ increases, whereas the profile depth increases. Therefore, a first shocking result of Figure 2 is that four MOVPE processes taking place at the same temperature (640 °C) for similar times –from 241 s in the black curve to 260 s in the magenta curve– yield remarkably different diffusion profiles. During the preexposure stage under study, the Ge substrates are submitted to an $AsH_3$ environment at high temperatures, which can etch their surfaces, modifying their morphology.[25] To quantify the surface roughening and investigate its correlation with the diverse As diffusion behaviors, another series of samples was prepared. New Ge substrates were exposed to different $t_{Preexp}$ followed by the growth of 4-5 nm GaAs layers to prevent oxide formation on the Ge surface, while preserving the underlying Ge surface roughness for the analysis.[26] In Figure 2(b), a dot graph shows the tendency of the surface RMS roughness when samples are exposed to different $t_{Preexp}$. The AFM scan images (Figure 2(c) and (d)) and the height profiles (Figure 2(e)) show also a peak to valley height significantly different between the samples using 1 and 20 second preexposure times.

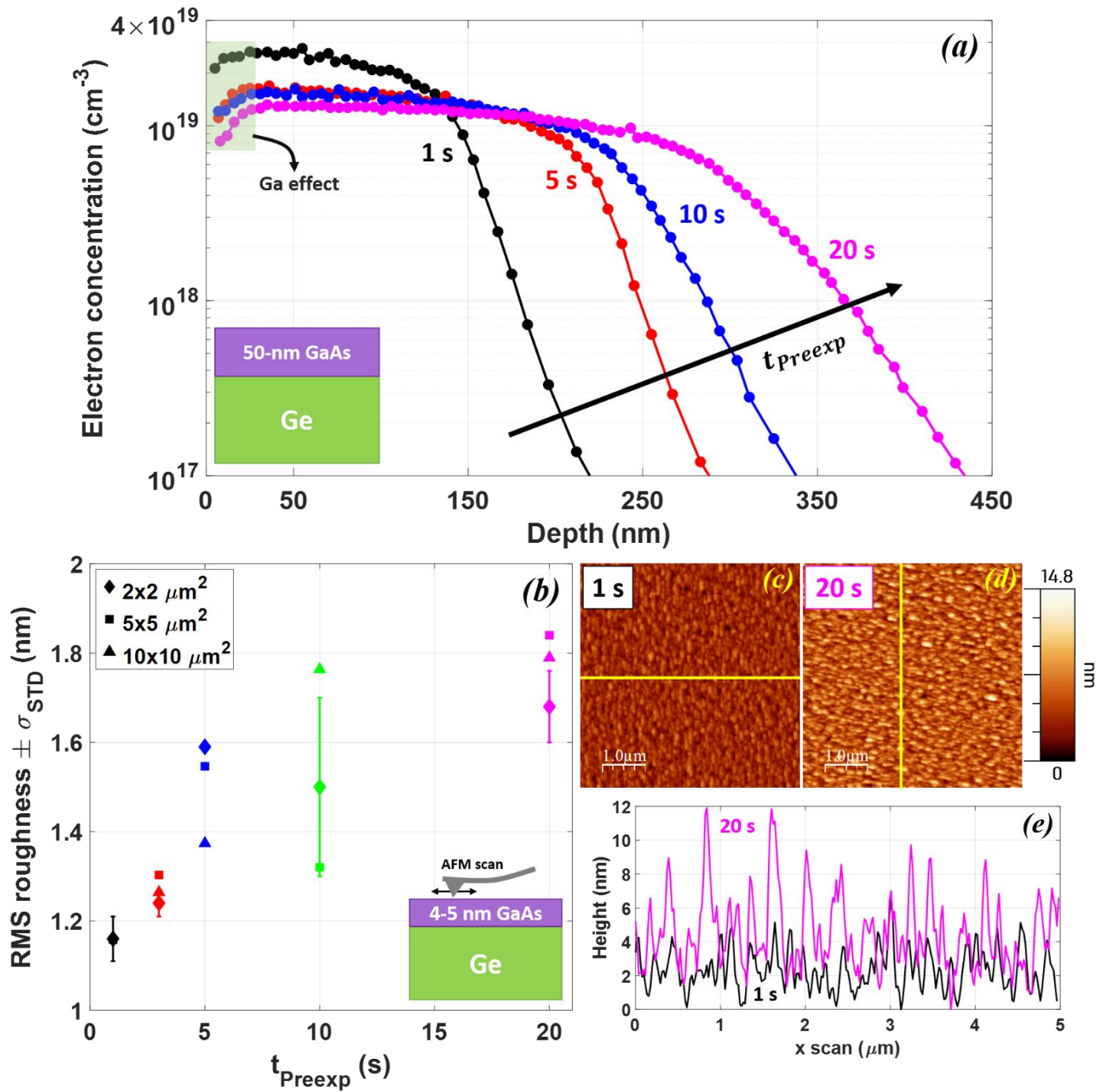

**Figure 2.** (a) ECV profiles for the samples with 50-nm GaAs and variable AsH$_3$ preexposure times (colored numbers). A simple illustration of the structure is placed in the lower left corner. (b) RMS roughness vs preexposure times. The RMS roughness obtained from different AFM scan areas is included (see legend). The RMS values for 2x2 µm$^2$ areas include standard deviation ($\sigma_{STD}$ ; represented by bars) whilst 5x5 µm$^2$ and 10x10 µm$^2$ scans were measured once. Two representative 5x5 µm$^2$ AFM scan images corresponding to 1-s (c) and 20-s (d) AsH$_3$ preexposure times are shown. (e) Height profiles represent the surface topography along the yellow lines overlapped on the AFM scan images above.

Once the impact of the AsH$_3$ preexposure has been described, the progress of the diffusion during different GaAs deposition times ($t_{GaAs}$) was examined. Our experimental design comprised two growth scenarios: (I) Samples without GaAs deposition or the omission of the stage 4 in Figure 1 and (II) samples with GaAs deposition, encompassing all stages. **Figure 3** compares two pairs of SIMS profiles, one from each of these growth scenarios. For all profiles, there is a high As concentration region with a steep decline before a 50 nm depth. After these regions, plateaus of As concentration can be seen, followed by a decrease towards the SIMS background noise. Samples subjected only to AsH$_3$ preexposure (triangles) do not exhibit significant differences, but SIMS profiles after GaAs depositions (circles) reveal that As diffusion is more intense for longer $t_{Preexp}$. Similar to the ECV measurements (Figure 2(a)), the 1-s profile (black circles) shows a diffusion depth which is roughly a half that on the 20-s profile (red circles).

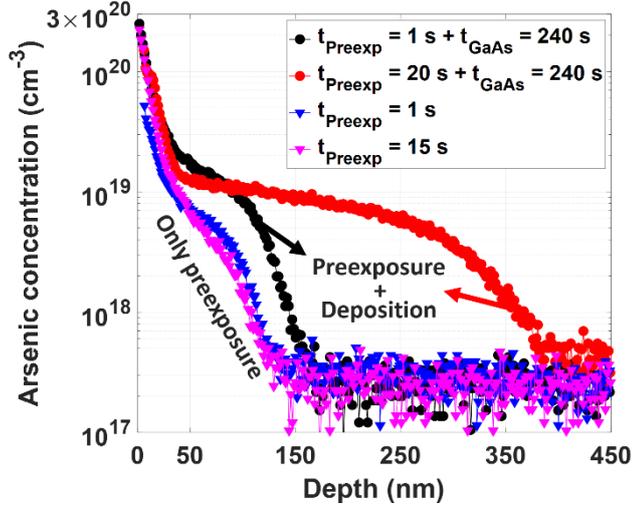 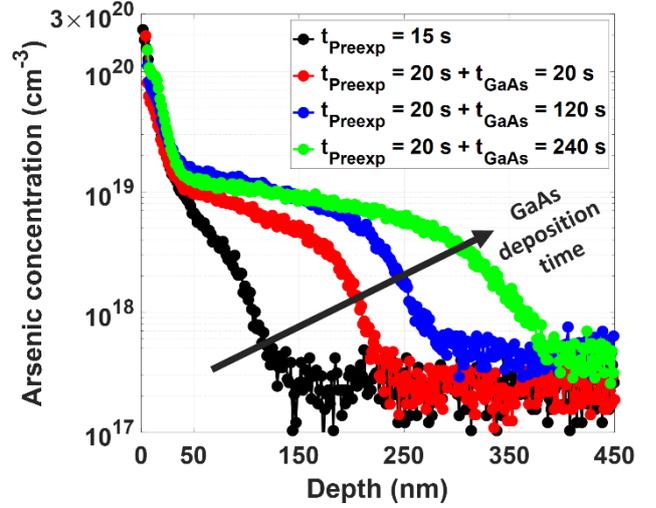

**Figure 3.** SIMS profiles of As for different growths. In downward triangles, profiles of samples exposed to AsH$_3$ for 1 s (blue) and 15 s (magenta). In filled circles, samples with 50-nm GaAs and different preexposure time of 1 s (black) and 20 s (red).

**Figure 4.** SIMS profiles of As for samples with different GaAs thicknesses and a constant preexposure time of 20 s. Profiles with 4-nm GaAs (red), 25-nm GaAs (blue) and 50-nm GaAs (green) are represented. Sample subjected to only AsH$_3$ preexposure for 15 s (black) is added to check directly the evolution of the As diffusion.

For the next experiment, we examine the influence of GaAs deposition using different GaAs deposition times (different thicknesses) while maintaining a constant $t_{Preexp}$= 20 s. The evolution of the As diffusion through several GaAs thicknesses is shown in **Figure 4**, where the SIMS profile of a sample with a similar preexposure and no GaAs layer deposited (black circles) is included in the graph for direct comparison. Unfortunately, the sample with exactly $t_{Preexp}$= 20 s was damaged during the first measurements and therefore we include a similar one with $t_{Preexp}$= 15 s. All profiles present an abrupt high concentration until approximately 50 nm, after which concentration plateaus gradually decline to finally drop at some depth into the Ge. It can be noticed that diffusion depths increase with growing GaAs thicknesses. As expected, longer deposition times result in higher arsenic doses, as evidenced by the areas under the profiles. This observation confirms the continued diffusion of As atoms into Ge during the GaAs deposition stage.

## 4. Discussion

As a basic tool to discuss the results, we present a model for As diffusion in Ge. This process has been described to be mediated by As-vacancy pairs (As$^+$V$^n$), which are formed by the combination of As$^+$ ions and charged vacancies in various charge states (V$^n$; n = 0, −1, −2).[10], [24] The effective diffusivity (D) based on this mechanism can be quantified as:

$$D(T, n) = D^0 + D^- \left(\frac{n}{n_i}\right) + D^= \left(\frac{n}{n_i}\right)^2 \qquad \text{Eq. (1)}$$

where D$^0$, D$^-$, and D$^=$ are the diffusivity coefficients for the three As-vacancy pairs, namely, As$^+$V$^0$, As$^+$V$^-$, and As$^+$V$^=$; $T$ is the temperature, $n$ is the free electron concentration, and $n_i$ the intrinsic carrier concentration in Ge. In turn, each of the three diffusion coefficients can be described by an Arrhenius equation with a specific activation energy ($E_A$) and pre-exponential diffusivity factor ($D_0^i$):

$$D^i(T) = D_0^i \exp\left(\frac{-E_A}{k_B T}\right) \qquad \text{Eq. (2)}$$

Many authors agree that double negatively charged vacancy pairing with an ionized As (As$^+$V$^=$) is the main mechanism for As diffusion in Ge and, therefore, **Eq. (1)** could be approximated with a quadratic dependence.[10], [11], [16], [24] However, this dependence stems from studies using processes different from MOVPE epitaxial growths, like ion implantation. In contrast to the ion-implanted As diffusion behavior observed in Ge, which demonstrates a quadratic dependence, our investigation conducted within a MOVPE environment utilizing AsH$_3$ indicates a linear dependency for arsenic diffusion. This observation aligns with previous findings reported by E. Takenaka et al., who investigated the behavior of arsenic diffusion in a MOVPE environment employing TBAs.[27] Following this precedent, Eq. (1) can be expressed as follows:

$$D(T,n) \cong D^-\left(\frac{n}{n_i}\right) = D_0^- \exp\left(\frac{-E_A}{k_B T}\right)\left(\frac{n}{n_i}\right) \quad \text{Eq. (3)}$$

where the activation energy and pre-exponential diffusivity factor which describe $D^-$ are $E_A = 1.9$ eV and $D_0^- = 6.2 \cdot 10^{-3}$ cm$^2 \cdot$s$^{-1}$.[27] This suggests that, under MOVPE conditions, at least when TBAs is the As precursor, diffusion is governed by pairs of a single negatively charged vacancy and an ionized As (As$^+$V$^-$), indicating a distinct mechanism at play.

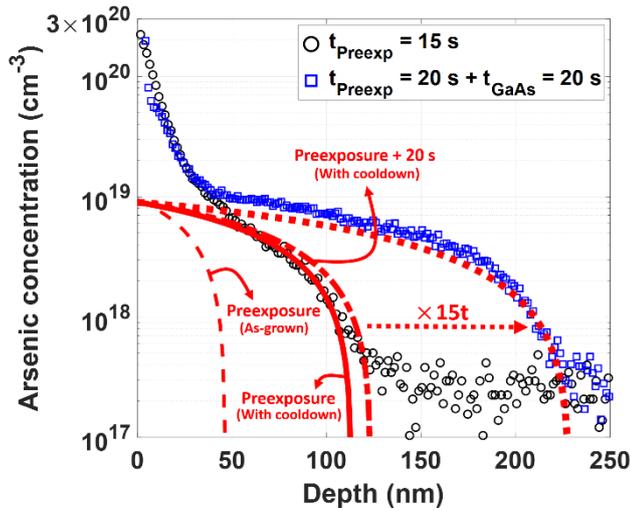
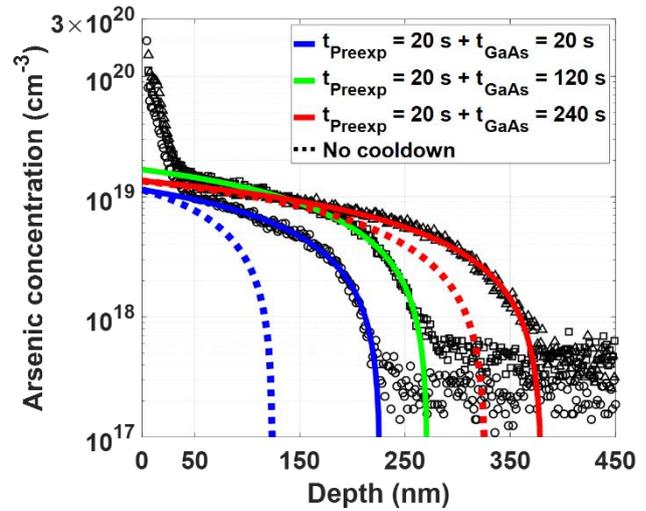

**Figure 5.** SIMS profiles (black and blue markers) and simulated profiles using the linear dependence described in Eq. (3) (red lines). In order of appearance, the thin red dashed line (As-grown) represents the resultant profile after a preexposure stage of 15 s with no cooldown. The same process but considering the cooldown is depicted in the red solid line, as well as a profile with 20 additional seconds (red dash-dotted line) beyond the red solid profile. The red dotted line represents another simulated profile that requires approximately 15 times of duration (from the red dash-dotted profile) to match the blue profile.

**Figure 6.** SIMS profiles for samples with identical preexposure times but different durations of GaAs deposition (black markers). The corresponding simulations are indicated (solid lines) with other simulations up to the start of the cooling down (dotted lines).

In order to get some quantitative insight into the diffusion process, we implemented a simulation approach based on **Eq. (3)**, namely, the model from Takenaka and coworkers.[27] Therefore, we considered that As diffusivity in Ge under MOVPE conditions follows a linear dependence on the free electron concentration. Diffusion profiles were simulated under this premise by utilizing the finite difference method under fixed surface concentrations ($C_0$), and with the activation energy given by Takenaka of 1.9 eV.[27] The first assumption is that every As atom is electrically active ($n$ = [As]), with other possible influences disregarded for simplicity. The simulation considers the temperature vs. time profile for each sample (Figure 1) and calculates the diffusion coefficient at each position in the structure as a function of the carrier concentration and the temperature. An important consideration to make is that this model only quantifies the diffusion process in the tail region and does not consider the subsurface region, where As solubility reaches the limit value.

**Figure 5** shows the SIMS profiles of a sample with only preexposure to AsH$_3$ (black markers), a sample with a similar preexposure followed by a short GaAs growth (blue squares) and their associated simulations (red lines). The red solid line represents the simulation of the sample with only preexposure to AsH$_3$, with fitting values of $C_0 = 9 \cdot 10^{18}$ cm$^{-3}$ and $D_0^- = 5.5 \cdot 10^{-3}$ cm$^2 \cdot$s$^{-1}$, respectively. This simulation follows quite reasonably the experiments in the tail region (black circles) with $C_0$ and $D_0^-$ values in close alignment with those already reported, where the parameters were determined by exposing Ge to TBAs (instead of AsH$_3$, as in this work).[27] Thus, it is reasonable to assume that Takenaka's model and associated parameters effectively characterize the behavior of the As diffusion during the preexposure stage also under AsH$_3$. In order to reproduce the evolution of the sample with preexposure and GaAs growth (blue squares), we applied this same model just increasing the diffusion time in 20 s (i.e. the GaAs growth time), obtaining the profile in the red dash-dotted line. However, the actual diffusion during these additional seconds (blue squares) results in As reaching much greater depths. To fit the experimental SIMS profile, using the previous model (with the same $C_0$ and $D_0^-$), the required time to match the blue curve would be 15 times longer (red dotted line). In other words, to accurately model the As diffusion process in MOVPE ambient we need to consider that GaAs deposition drastically increases the diffusivity of As. A final consideration around Figure 5 is related to the effect of the cooldown. The thin dashed red line in Figure 5 represents the As profile at the end of the preexposure, right before the moment when the heater is switched off and the cooldown starts (see Figure 1). So, the distance between the thin and thick red lines represents the As drive-in associated with the cooldown, which is indeed significant in our MOVPE reactor. This makes sense since the preexposure is just 20 s long, whereas, in the cooldown, it takes 5 min to go form the growth temperature of 640 ºC to 500 ºC where virtually the diffusion stops, according to our simulations.

In **Figure 6**, simulations were performed for samples with different GaAs deposition times ($t_{GaAs}$) but a constant $t_{Preexp}$ of 20 s. Three phases are considered in the simulation of diffusion profiles: 1) *preexposure*: the initial stage simulated using the parameters of Figure 5 (thin red dashed curve), identical for all samples; 2) *growth:* step with different duration depending on the sample; and 3) *cooldown*: the final stage when temperature steadily goes down but As still diffuses, identical for all samples. The colored thick lines in Figure 6 represent the profiles calculated using this approach. To obtain such good fits the average pre-exponential diffusivity factor ($D_0^-$) remains almost the same as for the preexposure-only samples, whilst $C_0$ needs to be increased ~150% on average. The diffusion model is simple so, rather than highlighting the particular values obtained for the parameters, our claim is that the same model used in Figure 5 (Eq. (3)), yields also reasonable fits for samples with GaAs growth provided that the diffusion parameters are increased at this stage, quantitatively supporting the enhancing role of GaAs epitaxial growth. Again, the dashed lines in Figure 6 represent the situation of the diffusion profile right before the onset of the cooldown (i.e. the as-grown profiles). As in the case of preexposure-only samples, the drive-in caused by the cooldown is notable, mostly for samples with short growth times.

The model just presented describes well some of the phenomena observed. When we just have Ge preexposure to AsH$_3$ in a MOVPE ambient, the diffusion behaves as reported by Takenaka et al. using TBAs, considering the effect of the preexposure plus the unavoidable thermal load of the cooldown, which plays a significant role because of the short process times. Therefore, for the preexposure times typically used (1 to 20 seconds), the resulting As profile is shallow and similar between samples. When growing GaAs, however, the diffusion of As accelerates. An enhanced dopant diffusion in Ge (as well as in Si) has been frequently linked to the generation of an extra concentration of point defects, particularly of vacancies in the saturated regions.[28]
In our case, we hypothesize that there are two key phases in the process, 1) during the preexposure, the chemical interaction of Ge from the wafer surface with foreign species coming from the gas phase can be a source of Ge vacancies; and 2) during epitaxial growth the wafer surface morphology will determine the properties of the GaAs/Ge interface and GaAs epilayer, governing the outdiffusion of Ge from the wafer into the layer,i.e. the injection of Ge vacancies). In addition, our high V/III ratio during GaAs deposition (~ 450) influences the injection of Ge vacancies. In As-rich GaAs epitaxy, the creation of gallium vacancies is favored, promoting the outdiffusion of Ge to the Ga lattice sites.[29] However, all samples in our study maintained equal V/III ratios and growth rates, suggesting that this phenomenon may not be a determining factor in explaining the variations observed in the As profiles.

Regarding the preexposure, significant chemical interaction is known to occur between AsH$_3$ (and its byproducts) and the Ge surface at our process temperatures.[25] Accordingly, the AFM measurements presented in Figure 2(b), (c), (d) and (e) have shown how AsH$_3$ affects the morphology of the wafer surface. The samples exposed to AsH$_3$ for longer times present a substantial increase in surface RMS roughness, resulting from Ge hydration and subsequent dimer displacement.[25] So, when AsH₃ is present in the reactor, the Ge wafer is etched, and a number of Ge atoms are removed from the pre-existing surface. We speculate that this creates a driving force for Ge atoms in the bulk to diffuse towards the surface to restore thermodynamic equilibrium (i.e. to minimize the surface energy by replenishing Ge atoms for the surface reconstruction) thus creating vacancies mostly in the subsurface region. In such process, most of the displaced Ge would form germane and desorb into the gas phase. In short, the AsH₃ preexposure phase represents a brief time in which vacancy injection takes place. Similar mechanisms have been suggested to explain phosphorus diffusion into Si when PH₃ is used as a precursor in a MOVPE ambient.[28]

During GaAs growth, the process that injects vacancies in the Ge wafer is the outdiffusion of Ge atoms, as the GaAs epilayer sucks in Ge atoms that leave a Ge vacancy behind. At a given temperature, the factors that limit the outdiffusion are the crystal orientation of the GaAs/Ge interface and the quality of the GaAs layer. When GaAs growth starts, the etching stops as AsH₃ is no longer in contact with Ge, and the surface morphology is frozen. If the preexposure has been short, etching would have been minimal, and the GaAs/Ge heterointerface will almost follow the nominal wafer orientation. This creates perfect conditions for the epitaxy and the crystalline quality of the epilayer is optimal. Contrarily, if the preexposure has taken longer, the wafer will roughen and thus start to exhibit (to some extent) higher index crystal planes, and, consequently, the GaAs/Ge heterointerface will present an equally rough morphology. This will impact the quality of the epitaxy yielding a layer with a higher concentration of crystal defects (point defects, antiphase domains, dislocations, …).

As evidenced by Figure 2, our results establish a clear correlation between the duration of AsH₃ preexposure and the resulting diffusion profile after GaAs deposition. However, all the curves in Figure 2(a) correspond to diffusion profiles that have been produced at analogous conditions, namely, the same temperature (640 ºC) and similar process times. For example, the black curve is the profile resulting from an AsH₃ preexposure of 1 s followed by 240 s of GaAs growth (i.e., 241 s total process time), whereas the magenta curve corresponds to an AsH₃ preexposure of 20 s followed by 240 s of GaAs growth (i.e., 260 s total process time). Our simulation shows that just a 19 s difference (~5%) in total process time cannot justify the blunt deviation of both diffusion profiles. Another key factor influencing the diffusivity is the initial free electron concentration ($n$), so it could be that differences in $n$ at the beginning of the diffusion process are causing the divergence amid the experiments in Figure 2. However, this is not the case either since extreme preexposure times yield very similar initial concentrations of As, as shown in Figure 3, where blue ($t_{Preexp}$= 1 s) and magenta ($t_{Preexp}$= 15 s) curves virtually overlap. In summary, neither the diffusion temperature, nor the duration of the diffusion process, nor the initial As⁺ concentration are substantially different in any of the experiments in Figure 2. Accordingly, it must be the concentration of vacancies in the Ge subsurface area, and its impact on the diffusion parameters, which should be the factor creating the notable differences amidst the diffusion profiles. In this regard, two different vacancy injection mechanisms determine the value of the diffusion parameters during preexposure and growth, respectively. We speculate that Ge outdiffusion is a stronger vacancy injection mechanism than AsH₃ etching and this explains why GaAs growth boosts diffusion. Even more so, Ge outdiffusion becomes more intense when it takes place across rough III-V/Ge heterointerfaces as it has been already reported.[30] Moreover, diffusion across high index planes becomes more intense as well.[27] These phenomena provide an explanation for the evolution of the profiles in Figure 2.

The hypothesis proposed provides a consistent explanation for the evolution of the profiles reported in our experiments and agrees with comparable data in the literature. Obviously, at this point, this is just a tentative explanation which needs further investigation –currently ongoing– to be substantiated.

## 5. Conclusions

High performance n/p junctions, the foundation of high-quality Ge-based photovoltaics, can be achieved by growth of GaAs layers on Ge with a proper control of the inevitable As diffusion into Ge. In this direction, the influence of MOVPE growth parameters on the diffusion of As in p-type Ge to form n/p junctions has been investigated through dedicated growth processes and fitting of the resulting As SIMS profiles by numerical simulations. Concerning the initial preexposure of the Ge wafer to AsH$_3$, the simulations demonstrate a linear dependence of the As diffusion coefficient with the electron concentration, in contrast to the quadratic dependence found in, for instance, ion-implanted As but in accordance with published results for diffusion in a MOVPE environment. We also see negligible As profile variations for the different AsH$_3$ preexposure times explored but an appreciable increase in the Ge surface roughness. The growth of a GaAs layer after the preexposure step is shown to enhance the diffusion of As, since the profiles simulated by just extending the diffusion time by the GaAs growth time result in notably shorter diffusion depths than in the experimental profiles. Moreover, for the same GaAs thickness deposited, the diffusion depth increases substantially for longer AsH$_3$ pre-exposures.

To analyze these findings, the nature of the As diffusion in Ge, which is mediated by Ge vacancy-As pairs is considered. The overabundance of As observed at the Ge surface can only be mobilized by injection of Ge vacancies. We hypothesize that the diffusion of As is enhanced during growth of GaAs by the injection of the vacancies left by Ge as it out-diffuses into the growing GaAs layer. Moreover, the effect of the pre-exposure stage can be explained also in terms of vacancy injection as the surface roughens, and by an enhanced Ge out-diffusion through the rough GaAs/Ge interface, as reported before. This increased availability of Ge vacancies is modelled through a higher surface concentration of As ready to diffuse, giving good fits to the experimental As profiles. These results demonstrate the intricate dependence of the As diffusion on the MOVPE process parameters and suggest the management of Ge vacancies injection as a way to gain control over the As profiles obtained.

## Acknowledgments

IGV acknowledges the support of project VIGNETRANSFER (grant number PDC2022-133445-I00) and IRS acknowledges the support of project TEIDE (grant number TED2021-131990B-I00), both funded by the Spanish Ministerio de Ciencia e Innovación/Agencia Estatal de Investigación [MCIN/AEI DOI: 10.13039/501100011033]. The epitaxy runs reported in this paper were carried out in a MOVPE reactor retrofitted with project RENOAIX200 (grant number EQC2019-005701-P) funded also by Spanish MCIN/AEI and FEDER "Una manera de hacer Europa". Part of the equipment used in this research for TPV cell manufacturing was acquired through project LABCELL30 [grant number EQC2021-006851-P] with funding from the Spanish Ministerio de Ciencia e Innovación /Agencia Estatal de Investigación [MCIN/AEI 10.13039/501100011033] and the European Union "NextGenerataionEU"/PRTR". V. Orejuela is funded by the Spanish Ministerio de Ciencia e Innovación (MCIN) through a FPI grant (PRE2019-088437).

## Conflict of interest

The authors declare no conflict of interest.

### References
[1] D. Ahn et al., 'High performance, waveguide integrated Ge photodetectors', *Opt. Express*, vol. 15, no. 7, Art. no. 7, 2007, doi: 10.1364/OE.15.003916.
[2] L. Vivien et al., 'Metal-semiconductor-metal Ge photodetectors integrated in silicon waveguides', *Applied Physics Letters*, vol. 92, no. 15, Art. no. 15, Apr. 2008, doi: 10.1063/1.2909590.
[3] V. Sorianello et al., 'Germanium on insulator near-infrared photodetectors fabricated by layer transfer', *Thin Solid Films*, vol. 518, no. 9, Art. no. 9, Feb. 2010, doi: 10.1016/j.tsf.2009.09.134.
[4] T. Burger, C. Sempere, B. Roy-Layinde, and A. Lenert, 'Present Efficiencies and Future Opportunities in Thermophotovoltaics', *Joule*, vol. 4, no. 8, Art. no. 8, Aug. 2020, doi: 10.1016/j.joule.2020.06.021.
[5] J. Fernández, F. Dimroth, E. Oliva, M. Hermle, and A. W. Bett, 'Back-surface Optimization of Germanium TPV Cells', in *AIP Conference Proceedings*, Madrid (Spain): AIP, 2007, pp. 190–197. doi: 10.1063/1.2711736.
[6] D. J. Friedman and J. M. Olson, 'Analysis of Ge junctions for GaInP/GaAs/Ge three-junction solar cells', *Prog. Photovolt: Res. Appl.*, vol. 9, no. 3, Art. no. 3, May 2001, doi: 10.1002/pip.365.

[7] B. Galiana, I. Rey-Stolle, C. Algora, K. Volz, and W. Stolz, 'A GaAs metalorganic vapor phase epitaxy growth process to reduce Ge outdiffusion from the Ge substrate', *Appl. Phys. Lett.*, vol. 92, no. 15, p. 152102, Apr. 2008, doi: 10.1063/1.2901029.

[8] L. Barrutia, I. García, E. Barrigón, M. Ochoa, C. Algora, and I. Rey-Stolle, 'Impact of the III-V/Ge nucleation routine on the performance of high efficiency multijunction solar cells', *Solar Energy Materials and Solar Cells*, vol. 207, p. 110355, Apr. 2020, doi: 10.1016/j.solmat.2019.110355.

[9] W. He *et al.*, 'Structural and optical properties of GaInP grown on germanium by metal-organic chemical vapor deposition', *Appl. Phys. Lett.*, vol. 97, no. 12, Art. no. 12, Sep. 2010, doi: 10.1063/1.3492854.

[10] J. Vanhellemont and E. Simoen, 'On the diffusion and activation of n-type dopants in Ge', *Materials Science in Semiconductor Processing*, vol. 15, no. 6, Art. no. 6, Dec. 2012, doi: 10.1016/j.mssp.2012.06.014.

[11] A. Chroneos and H. Bracht, 'Diffusion of *n* -type dopants in germanium', *Applied Physics Reviews*, vol. 1, no. 1, Art. no. 1, Mar. 2014, doi: 10.1063/1.4838215.

[12] P. Martín, V. Orejuela, C. Sanchez-Perez, I. García, and I. Rey-Stolle, 'Device Architectures for Germanium TPV Cells with Efficiencies over 30%', in *2023 14th Spanish Conference on Electron Devices (CDE)*, Valencia, Spain: IEEE, Jun. 2023, pp. 1–4. doi: 10.1109/CDE58627.2023.10339503.

[13] S. Koffel *et al.*, 'Experiments and simulation of the diffusion and activation of the n-type dopants P, As, and Sb implanted into germanium', *Microelectronic Engineering*, vol. 88, no. 4, Art. no. 4, Apr. 2011, doi: 10.1016/j.mee.2010.09.023.

[14] K. Benourhazi and J. P. Ponpon, 'Implantation of phosphorous and arsenic ions in germanium', *Nuclear Instruments and Methods in Physics Research Section B: Beam Interactions with Materials and Atoms*, vol. 71, no. 4, Art. no. 4, Sep. 1992, doi: 10.1016/0168-583X(92)95358-X.

[15] H. Bracht and S. Brotzmann, 'Atomic transport in germanium and the mechanism of arsenic diffusion', *Materials Science in Semiconductor Processing*, vol. 9, no. 4–5, Art. no. 4–5, Aug. 2006, doi: 10.1016/j.mssp.2006.08.041.

[16] S. Brotzmann and H. Bracht, 'Intrinsic and extrinsic diffusion of phosphorus, arsenic, and antimony in germanium', *Journal of Applied Physics*, vol. 103, no. 3, Art. no. 3, Feb. 2008, doi: 10.1063/1.2837103.

[17] C. K. Chia *et al.*, 'Effects of AlAs interfacial layer on material and optical properties of GaAs∕Ge(100) epitaxy', *Appl. Phys. Lett.*, vol. 92, no. 14, Art. no. 14, Apr. 2008, doi: 10.1063/1.2908042.

[18] T. T. Chiang and W. E. Spicer, 'Arsenic on GaAs: Fermi-level pinning and thermal desorption studies', *Journal of Vacuum Science & Technology A*, vol. 7, no. 3, Art. no. 3, May 1989, doi: 10.1116/1.575874.

[19] E. Barrigon, B. Galiana, and I. Rey-Stolle, 'Reflectance anisotropy spectroscopy assessment of the MOVPE nucleation of GaInP on germanium (1 0 0)', *Journal of Crystal Growth*, vol. 315, no. 1, Art. no. 1, Jan. 2011, doi: 10.1016/j.jcrysgro.2010.09.038.

[20] S. Gan, L. Li, M. J. Begarney, D. Law, B.-K. Han, and R. F. Hicks, 'Step structure of arsenic-terminated vicinal Ge (100)', *Journal of Applied Physics*, vol. 85, no. 3, Art. no. 3, Feb. 1999, doi: 10.1063/1.369176.

[21] H. Kroemer, 'Polar-on-nonpolar epitaxy', *Journal of Crystal Growth*, vol. 81, no. 1, Art. no. 1, Feb. 1987, doi: 10.1016/0022-0248(87)90391-5.

[22] E. Barrigón, S. Brückner, O. Supplie, H. Döscher, I. Rey-Stolle, and T. Hannappel, 'In situ study of Ge(100) surfaces with tertiarybutylphosphine supply in vapor phase epitaxy ambient', *Journal of Crystal Growth*, vol. 370, pp. 173–176, May 2013, doi: 10.1016/j.jcrysgro.2012.07.046.

[23] A. Paszuk *et al.*, 'Atomic surface control of Ge(100) in MOCVD reactors coated with (Ga)As residuals', *Applied Surface Science*, vol. 565, p. 150513, Nov. 2021, doi: 10.1016/j.apsusc.2021.150513.

[24] C. L. Claeys and E. Simoen, Eds., *Germanium-based technologies: from materials to devices*. Amsterdam ; Boston: Elsevier, 2007.

[25] W. E. McMahon and J. M. Olson, 'Atomic-resolution STM study of a structural phase transition of steps on vicinal As/Ge(100)', *Phys. Rev. B*, vol. 60, no. 23, pp. 15999–16005, Dec. 1999, doi: 10.1103/PhysRevB.60.15999.

[26]   J. Oh and J. C. Campbell, 'Thermal desorption of Ge native oxides and the loss of Ge from the surface', *Journal of Elec Materi*, vol. 33, no. 4, pp. 364–367, Apr. 2004, doi: 10.1007/s11664-004-0144-4.

[27]   M. Takenaka, K. Morii, M. Sugiyama, Y. Nakano, and S. Takagi, 'Gas Phase Doping of Arsenic into (100), (110), and (111) Germanium Substrates Using a Metal–Organic Source', *Jpn. J. Appl. Phys.*, vol. 50, no. 1R, p. 010105, Jan. 2011, doi: 10.1143/JJAP.50.010105.

[28]   E. García-Tabarés, D. Martín, I. García, and I. Rey-Stolle, 'Understanding phosphorus diffusion into silicon in a MOVPE environment for III–V on silicon solar cells', *Solar Energy Materials and Solar Cells*, vol. 116, pp. 61–67, Sep. 2013, doi: 10.1016/j.solmat.2013.04.003.

[29]   G. Timo *et al.*, 'The effect of the growth rate on the low pressure metalorganic vapour phase epitaxy of GaAs/Ge heterostructures', *Journal of crystal growth*, vol. 125, no. 3, pp. 440–448, 1992.

[30]   P. J. Sophia *et al.*, 'Influence of Surface Roughness on Interdiffusion Processes in InGaP/Ge Heteroepitaxial Thin Films', *ECS J. Solid State Sci. Technol.*, vol. 4, no. 3, pp. P53–P56, 2015, doi: 10.1149/2.0021503jss.